\documentclass[12pt]{article}

\pdfoutput=1

\usepackage[utf8]{inputenc}
\usepackage[DIV=13]{typearea}
\usepackage{ragged2e}
\usepackage{amsmath,amsfonts,bbm,mathrsfs,amssymb}
\usepackage{slashed,cancel}
\usepackage[usenames,dvipsnames]{xcolor}
\usepackage{cite}
\usepackage{bm} 
\usepackage{hyperref}
\hypersetup{colorlinks=true,urlcolor=blue,anchorcolor=blue,citecolor=blue,filecolor=blue,
            linkcolor=blue,menucolor=blue, linktocpage=true,pdfproducer=medialab}
\usepackage[english]{babel}
\usepackage{catchfile}
\usepackage{indentfirst}
\usepackage[scriptsize]{subfigure}
\usepackage{graphicx}
\usepackage{float}
\usepackage{comment}
\usepackage{enumerate}
\usepackage{multirow}
\usepackage[normalem]{ulem} 
\usepackage{ytableau}
\usepackage{mathtools}
\usepackage{xcolor}

\def\eq#1{{Eq.~(\ref{#1})}}

\newcommand{\be}{\begin{equation}}
\newcommand{\ee}{\end{equation}}
\newcommand{\ba} {\begin{equation}\begin{aligned}}
\newcommand{\ea} {\end{aligned}\end{equation}}
\newcommand{\bg} {\begin{equation}\begin{gathered}}
\newcommand{\eg} {\end{gathered}\end{equation}}

\newcommand*{\INFNFR}{Istituto Nazionale di Fisica Nucleare, Laboratori Nazionali di Frascati, C.P. 13, 00044 Frascati, Italy}
\newcommand*{\LNGS}{Istituto Nazionale di Fisica Nucleare, Laboratori Nazionali del Gran Sasso, Assergi, 67100, Italy}
\newcommand*{\NICPB}{Laboratory of High Energy and Computational Physics, NICPB, R\"avala 10, 10143, Tallin, Estonia}
\newcommand*{\ENS}{Département de Physique, École Normale Supérieure de Lyon,  46 Allée d'Italie,
F 69342, Cedex 07 Lyon, France}

\title{\textbf{Atoms as Electron Accelerators \\ for New Physics Searches}}

 \author{Fernando Arias-Aragón,$^{1}$\footnote{ \url{fernando.ariasaragon@lnf.infn.it}} \,
 Giovanni Grilli di Cortona,$^{2}$\footnote{ \url{giovanni.grilli@lngs.infn.it}}\\  Enrico Nardi,$^{1,3}$\footnote{ \url{enrico.nardi@lnf.infn.it}}\, Léo Veissière$^{4}$\footnote{ \url{leo.veissiere@ens-lyon.fr}}}
 \vspace{2cm}	
\date{ \footnotesize
    $^1$  \textit{\INFNFR} \\
    $^2$  \textit{\LNGS }  \\
    $^3$  \textit{\NICPB } \\
    $^4$  \textit{\ENS }       
	}

\begin{document}
\maketitle

\begin{abstract}
Due to Heisenberg's uncertainty principle, atomic electrons localized around the nucleus exhibit a characteristic momentum distribution that, in elements with high atomic number, remains significant up to  relativistic values.
Consequently, in fixed-target experiments, atoms can effectively act 
as electron accelerators, increasing the centre-of-mass energy 
in collisions with beam particles.
In this work, we leverage this effect to explore its potential for new physics searches. We consider positrons from beams of various energies annihilating with 
atomic electrons in  a $^{74}$W fixed target. We compute the production rates of new vector bosons and pseudoscalar particles as functions of their couplings and masses.
We show that the electron-at-rest approximation significantly underestimates the mass reach for producing these new states compared to the results obtained by properly accounting for atomic electron momenta.
In particular, we estimate the sensitivity for detecting these new particles using the positron beam at the Beam Test Facility linac at the Laboratori Nazionali di Frascati, the  H4 beamline in the CERN North Area, and the proposed Continuous Electron Beam Accelerator Facility 
of  Jefferson Laboratory.
\end{abstract}

\newpage
\tableofcontents

\section{Introduction}

In the quest for physics Beyond the Standard Model (BSM), several experimental strategies are employed, ranging from astrophysical and cosmological observations to Earth-based laboratory experiments. Regarding accelerator experiments, two main setups exist. On the one hand, colliders, both circular and linear, accelerate two beams of particles and direct them into head-on collisions. On the other hand, in 
fixed-target experiments a single beam of particles is accelerated and  directed at a target material.  Both approaches have their own advantages and disadvantages.
Colliders can fully exploit the energy of the accelerated beams 
to produce known or new particles of higher mass, 
as the centre-of-mass (c.m.) of the collision 
is at rest in the laboratory frame, at least in first approximation. 
In contrast, in fixed-target experiments, a large fraction of the available energy 
is carried away by the forward boost of the produced particles,  limiting the accessible 
 mass range for new particle production.
 However, a key advantage of this approach is the significantly higher particle density in targets or dumps, many orders of magnitude greater than in a colliding beam. This makes such experiments a powerful tool for probing feebly interacting particles~\cite{Antel:2023hkf}, which are characterized by sub-electroweak scale masses and extremely small couplings to the Standard Model (SM) sector, despite their highly suppressed production cross-section.

Recently, it was realised that the drawback of a reduced c.m. energy 
in fixed target experiments can be partially compensated by leveraging the 
intrinsic velocities that characterise  bound  electrons localised 
around the nucleus~\cite{Arias-Aragon:2024qji}. The effects of electron motion  in positron annihilation off fixed-target atomic electrons were first discussed in Ref.~\cite{Nardi:2018cxi}, where they were identified as a nuisance 
in the search for a hypothetical, very narrow resonance
commonly denoted as  $X_{17}$ due to its approximate mass of 17\,MeV.\footnote{The reason why electron motion reduces the sensitivity of  $X_{17}$ searches via 
$e^+e^-$  resonant annihilation is that the electron momentum distribution 
causes a spread in the c.m. energy of the  collision.  
This renders precise tuning 
of the beam energy ineffective, as well as any effort to reduce the beam energy spread.} 
This resonance  was proposed to 
explain certain anomalies observed in $e^+e^-$ internal pair creation in nuclear transitions~\cite{Krasznahorkay:2015iga,Krasznahorkay:2021joi,
Krasznahorkay:2022pxs,Krasznahorky:2024adr}, sparking significant interest  
in the community (see Ref.~\cite{Alves:2023ree} for a review).
However, in Ref.~\cite{Arias-Aragon:2024qji}, it was recognized that, especially in high-$Z$ targets, the c.m. energy in electron-positron head-on collisions can increase by a large factor due to the electron motion. This phenomenon effectively enhances the mass reach of searches for BSM particles compared to previous estimates based on the free-electron-at-rest approximation.

In this work, we study BSM particle searches via their resonant production 
in positron annihilation on atomic electrons, properly accounting for 
electron momentum effects. We consider three different positron beamlines: 
the Beam Test Facility (BTF) linac~\cite{Buonomo:2023pzi}  at the Laboratori Nazionali di Frascati (LNF) 
serving the Positron Annihilation into Dark Matter 
Experiment (PADME)~\cite{Raggi:2014zpa}; the proposed positron beam 
at the Continuous Electron Beam Accelerator Facility (Ce$^+$BAF) at Jefferson Laboratory (JLab)~\cite{Accardi:2020swt,
Grames:2023pre} and the 
 H4 beamline~\cite{Gatignon:2730780} 
at the CERN North Experimental Area~\cite{Banerjee:2774716}.
In all cases, we will estimate the production rates of new BSM particles on thick targets, simultaneously accounting for positron energy loss while traversing the target and atomic electron motion.

Searches for displaced decays into SM particles, using similar experimental setups (commonly referred to as beam dump experiments), have historically yielded some of the most sensitive direct BSM searches, benefiting from large luminosities and the boosted kinematics of the final states~\cite{Riordan:1987aw,Bross:1989mp,Konaka:1986cb,Bjorken:1988as,Davier:1989wz,Bergsma:1985qz,Barabash:1992uq}.
In the broader experimental effort to detect feebly interacting particles, these searches complement 
$e^+e^-$ 
  colliders and fixed-target missing energy/momentum experiments, offering strong sensitivity to light BSM particles with masses up to 
  several hundred MeV~\cite{Antel:2023hkf}.  Remarkably, a region of parameter space remains unconstrained between the beam dump and collider sensitivities. This corresponds to a range of couplings that are too small to ensure sufficiently large production rates at colliders, yet too large to allow BSM particle to decay outside the fixed targets.
Closing this gap is challenging and has been a key focus of experimental efforts in recent years.~\cite{Antel:2023hkf}.

In this work, we demonstrate that using positron beams can improve BSM particle production rates by orders of magnitude due to resonant production, while atomic electron motion further extends the experimental mass reach. The combination of both effects would enable an experiment with advanced vertexing capabilities to probe unexplored regions within the sensitivity gap. 
We will discuss and present results for two types of BSM particles: vector and pseudoscalar bosons, for which there are sound theoretical motivations for relatively light masses. However, we have verified 
that to an excellent approximation, the exclusion region for 
axial-vector (scalar) particles coincides with that of vector (pseudoscalar).

 To incorporate the effects of the electron momentum distribution
in evaluating the vector and pseudoscalar production rates, a target 
material must be specified.
We will present results for positron annihilation on a $^{74}$W target,
as tungsten is commonly used as target material. 
However, it should be stressed that other higher-$Z$ materials, such as $^{84}$Pb (despite potential concerns regarding its low melting temperature), and up to $^{90}$Th or $^{92}$U (despite possible issues with radio-activation), should be considered interesting alternative options to explore, given the significant enhancement of electron momentum effects compared to $^{74}$W.

\section{Resonant production}

The production of generic dark sector particles $P$ via resonant  electron-positron annihilation in fixed-target experiments can achieve a significantly extended reach in the dark particle mass $m_P$ when high-$Z$ targets are used~\cite{Arias-Aragon:2024qji}. This  arises because the motion of atomic electrons can effectively increase the c.m. energy of the process. Incorporating atomic electron motion, 
the production cross section can be written in closed form as:
\bg
\sigma = \int_{k_A^{\mathrm{min}}}^{k_A^\mathrm{max}}dk_A\frac{\left|\mathcal{M}_{free}\right|^2k_A n(k_A) }{16\pi p_B|E_B k_A x_0(k_A)-E_{k_A} p_B|},\vspace{.2cm}\\ 
x_0(k_A) = \frac{2E_A E_B+2m_e^2-m_P^2-k_A^2 }{2k_Ap_B},\vspace{.2cm}\\
 k_A^\mathrm{max,min} = \left|p_B \pm \sqrt{ \left(E_A + E_B \right)^2-m_P^2} \right|\ ,
\label{eq:ResCS}
\eg
where the indices A and B refer to  atomic electrons and to  beam positrons, respectively.  The electron (positron) momentum and energy are denoted as 
 $k_A$ ($p_B$) and $E_A$ ($E_B$), while $E_{k_A}=\sqrt{k_A^2+m_e^2}$ is the free electron dispersion relation, as opposed to $E_A=m_e+u_q$ which refers to the energy of the bound electron, with $u_q$ the binding energy.
 Throughout this work we will neglect binding energy effects and take
 $E_A\simeq m_e$. For the processes we  focus on, this is a safe approximation.   
 However, for other processes, binding energy effects can be significant and must be 
 accounted for~\cite{Plestid:2024xzh,Plestid:2024jqm}. 
 Finally, $n(k_A)$ represents the electron momentum density function, which can be  directly related to the Compton profile~\cite{WilliamsBG1977,Arias-Aragon:2024qji}. 
 Specifically, we used Ref.~\cite{BIGGS1975201} for the theoretical Compton profile of tungsten in the range $0 \leq k_A < 100\,\textrm{a.u.} \simeq 370\,\mathrm{keV}$. For higher momenta  we have complemented  the 
 Compton profile calculation by   numerically estimating the contribution from the core orbitals up to the $3d$ shell, using the DBSR-HF code~\cite{ZATSARINNY2016287}. 

Most fixed-target experiments are performed with thick-targets, 
that is with targets of  length $z_{T}$  of several radiation length, typically of a few cm in the case of tungsten. 
Accordingly, we will consider  a 5 cm tungsten target for all the three
experimental setups discussed (LNF, JLab and CERN).
With the relatively low-energy BTF beam at LNF, such thick-target will absorb a large component of the background from electromagnetic showers generated within the target. 
For the higher-energy beams at JLab and CERN this shielding effect will be much less effective. Clearly, a realistic assessment of the effective sensitivity of each experiment would require a dedicated simulation of the detector and background, which is beyond the scope of this work.

In traversing a thick target, primary positrons continuously lose part of their energy, primarily through photon bremsstrahlung. As detailed in  Ref.~\cite{Nardi:2018cxi},  this effect can be leveraged  to  
continuously scan the c.m. energy in search of the $X_{17}$ boson  
using a positron beam with energy exceeding the resonant energy.
More recently it was also shown that harnessing the 
momentum distribution of  atomic target electrons enables scanning  the 
energy dependence of the hadronic cross 
section $ \sigma(e^+e^- \to\, {\rm hadrons})$  
from the two pion threshold up to $\sqrt{s} \sim 1\,$GeV, with a 12\,GeV positron 
beam~\cite{Arias-Aragon:2024gpm}. Note that, with electrons at rest, such a beam energy would not even be sufficient to reach the threshold for di-muon production.
In this study, we demonstrate how the combination of these two effects allows for the probing of a remarkably wide range of BSM particle masses through their resonant production, all while keeping the beam energy constant.
We assume that the newly produced particle $P$ decays into a pair of charged 
leptons, that are subsequently detected. It is therefore essential to consider the target length and properly account for the probability of $P$ decaying after exiting  the target, as well as the target-detector distance, and require that $P$ decays upstream of the detector. Accounting for both conditions, the expected number $N_P$ of $e^+e^- \to P \to \ell^+ \ell^-$ events (with $\ell=e,\,\mu$)   
can be written as:
\ba\label{eq:NEvents}
N_P &= \frac{N_{\mathrm{e}^+oT}N_{Av}Z\rho X_0}{A}\\
&\times \int_0^{z_{T}/X_0} dt\int dE_e \ \left(1-e^{\frac{z_{T}-z_{D}}{\ell_P}}\right) e^{\frac{X_0 t-z_{T}}{\ell_P}} \int dE\ \mathcal{G}\left(E,E_B,\sigma_B\right)I\left(E,E_e,t\right)\sigma\left(E_e\right),
\ea
where $N_{Av}$ is Avogadro's number,  $z_{T}$ and $z_{D}$ denote respectively the target length and the distance between the beginning of the target and the detector, and $\ell_P$ denotes the $P$ decay length given by
\be
\label{eq:boost}
\ell_P = \frac{\gamma_P}{\Gamma_P} \simeq
\frac{m_e+E_e}{m_P\Gamma_P},
\ee
with $\gamma_P =
 \left(1 + \frac{\vec p_P^{\;2}}{m_P^2}\right)^{1/2}
\simeq \frac{m_e+E_e}{m_P}$
the particle boost factor  and $\Gamma_P$ its total decay width. 
The target is characterized by the atomic number $Z$, mass number $A$, density $\rho$ and radiation length $X_0$.

The positron beam is assumed to have a Gaussian energy distribution $\mathcal{G}\left(E,E_B,\sigma_B\right)$, centred at the nominal beam energy $E_B$ with a standard deviation of $\sigma_B$, while $N_{\mathrm{e}^+oT}$ denotes the number of positrons on target. Positron energy loss within the target is described by the function
\be
I\left(E,E_e,t\right)= \frac{\Theta\left(E-E_e\right)}{E\ \Gamma(bt)}\left(\log{\frac{E}{E_e}}\right)^{bt-1},
\ee
which represents the probability that a positron entering the target with energy $E$ will have energy $E_e$ after traversing $t$ radiation lengths~\cite{Bethe:1934za,Tsai:1966js}. In the exponent, $b=4/3$, while $\Gamma$ and $\Theta$ denote respectively the Gamma function and the Heaviside step function.

\section{Dark sector models}

In this section, we briefly outline the extensions of the SM 
on which we will focus, and we collect the relevant formulae describing 
resonant particle production and decays. 
The BSM particles that we will consider  are  new vector bosons $X_\mu$ and pseudoscalar particles $a$.
For simplicity, we assume that these dark particles couple only to the SM charged leptons. 
Relaxing this assumption would not affect the production rates in the three experimental setups considered. 
However, possible couplings to lighter dark sector particles and, in the  range of higher vector and pseudoscalar masses, to SM hadrons, could lead to additional decay channels. 
This would shorten the particle's decay length for a given coupling to leptons, thereby affecting the detection characteristics of the charged dilepton signal (see below). 
We will also restrict our analysis to tree-level decays and therefore will not consider 
decays into two-photon, which, although loop-suppressed, are inevitably induced via charged lepton loops.

A vector particle $X_\mu$ interacting with $e^\pm$  may emerge as a gauge boson in theories beyond the SM where a flavour subgroup has been gauged, as in the $B-L$~\cite{Mohapatra:1980qe,Khalil:2006yi}, $L_e-L_\mu$ or 
 $L_e-L_\tau$~\cite{Foot:1990mn,He:1991qd,Foot:1994vd} models, or in models where a dark photon (DP) emerges from a generic new $U(1)$ gauge symmetry. 
The Lagrangian for a  vector particle with mass $m_X$ interacting with  SM leptons  $\ell$ reads
\be\label{eq:Vec}
\mathscr{L}_{X} \supset -\frac{1}{4} X_{\mu\nu}X^{\mu\nu}+ \frac{1}{2}m_X^2 X_\mu X^\mu + \sum_\ell g_{V\ell} X_\mu \bar{\ell} \gamma^\mu \ell,
\ee  
where $X_{\mu\nu} = \partial_\mu X_\nu - \partial_\nu X_\mu$ is the field strength of the vector $X_\mu$ and $g_{V\ell}$ is the coupling between $X$ and the lepton $\ell$. In the case of a DP, $g_{V\ell}$ can be related to the kinetic mixing parameter $\epsilon$ as $\epsilon = g_{V\ell}/e$, with $e$ denoting the electron charge~\cite{Holdom:1985ag}. This relation arises from the renormalizable and gauge-invariant term:  
\be
\mathscr{L}_{X} \supset  -\frac{\varepsilon}{2} X_{\mu\nu} F^{\mu\nu},
\ee  
where $F_{\mu\nu} $  is the electromagnetic field strength tensor.  

Following Ref.~\cite{Arias-Aragon:2024qji}, we write 
the matrix element for the resonant production 
process $e^+(P_B) + e^-(K_A) \to X(P_X)$,
averaged over initial spins and summed over final polarizations, in terms of products 
of the particles four-momenta: 
\be
|{\mathcal{M}_{free}}|^2 = g_V^2\left[3 m_e^2 + K_A\cdot P_B + \frac{2}{m_X^2} (K_A\cdot P_X) (P_B\cdot P_X)\right], 
\label{eq:Mvec}
\ee
where  $P_B$ and $P_X$ are the usual four-momenta of free on-shell particles,  
while $K_A$ has time-like component $E_{k_A}=\sqrt{k_A^2 +m_e^2}$, which 
differs from the energy of the bound electron $E_A\simeq m_e$
(hence the label $free$ for the matrix element, see Ref.~\cite{Arias-Aragon:2024qji}). 
The decay width of the vector particle to  SM leptons is given by
\be
\Gamma_{X\ell\ell} = \frac{m_X}{12\pi} \sum_\ell g_{V\ell}^2\left(1+2\frac{m^2_\ell}{m^2_X}\right)\sqrt{1-4\frac{m_\ell^2}{m_X^2}}\,\Theta(m_X-2m_\ell),
\ee
where the sum extends over $\ell=e,\,\mu,\,\tau$ and $m_\ell$ denotes the lepton mass.

Light pseudoscalar particles are naturally produced from the spontaneous breaking of a global symmetry when any source of explicit breaking remains sufficiently suppressed, and are also commonly denoted as pseudo Nambu-Goldstone bosons or 
axion like particles. 
We will only consider pseudoscalar interactions with leptons  
after electroweak symmetry breaking, which are described by the effective operator
\be\label{eq:Pseudo}
\mathscr{L}_{a}\supset  i\, \sum_\ell\, m_\ell\, g_{a\ell}\, a\,\bar{\ell}\gamma_5 \ell,
\ee
where $a$ is the pseudoscalar field and $g_{a\ell}$ is the pseudoscalar coupling to the specific lepton of mass $m_\ell$. We  assume that the couplings $g_{a\ell}$ and the pseudoscalar mass $m_a$ are independent parameters, as is characteristic of an axion-like particle.
The matrix element for  resonant production of a pseudoscalar $e^+(P_B)+e^-(K_A) \to a(P_a)$ averaged over initial spins is
\be
|{\mathcal{M}_{free}}|^2 = g_{a\ell}^2\,m_e^2\,(K_A\cdot P_B + m_e^2), 
\ee
where the notation is the same as in Eq.~\ref{eq:Mvec}.
The decay width of a  pseudoscalar particle to  SM leptons is given by 
\be
\Gamma_{a\ell\ell} = \frac{1}{8\pi} \sum_\ell g_{a\ell}^2 m_\ell^2\,\sqrt{m_a^2-4\,m_\ell^2}\,\Theta(m_a-2m_\ell).
\ee

\section{Results}

In this section, we present results on the achievable reach in mass and couplings for new vector bosons and pseudoscalars using positron beams 
from three accelerator facilities: the  
Frascati BTF linac~\cite{Buonomo:2023pzi}; 
the proposed Ce$^+$BAF at JLab~\cite{Accardi:2020swt,
Grames:2023pre} and the 
 CERN H4 beamline~\cite{Gatignon:2730780}.
The BTF linac can produce positron beams with a nominal energy $E_B$ within the range $250\leq E_B/\mathrm{MeV} \leq 450$. The maximum  
 available intensity is set by the radioprotection limit at 
$10^{18}$ e$^+$oT per year.\footnote{Technically, BTF can deliver up to $10^{20}$ e$^+$oT per year.}
The Ce$^+$BAF injector at JLab is planned to deliver extremely  
intense  positron beams. For unpolarised beams, achieving currents 
between $1$ and $5\,\mu$A and  energies of up to $12\,$GeV is challenging, but not unrealistic~\cite{Voutier}.  In our study we take a beam energy $E_B = 12\,$GeV and a 
total of $10^{21}$ e$^+$oT, corresponding to one year of data taking with a $5\,\mu$A current. 
The beamlines in the CERN North Experimental Area can deliver positrons with energies an order of magnitude higher, making it natural to explore how this tenfold increase would impact the search for BSM particles.
However, a major drawback is that these positron beams are tertiary, resulting in reduced intensity and significant hadron contamination at beam energies  above
100\,GeV.
Spills of $400\,$GeV protons from the SPS first strike a beryllium target, generating a wide range of particles. Charged particles are deflected, while secondary photons and photons from $\pi_0$ decays produce $e^+e^-$ pairs in a downstream lead converter. Magnetic fields and collimators are subsequently employed to select particles based on charge and momentum. A representative rate for the CERN H4 beamline, which serves the NA64 experiment, is $5 \times 10^6 \, e^+$oT per spill. Assuming 3500 spills per day, this corresponds to approximately $6.5 \times 10^{12}$ e$^+$oT per year. 
Recently, the feasibility of delivering a high-energy, high-intensity positron beam to CERN's ECN3 cavern via the K12 beamline, which serves the NA62 experiment, has also been explored~\cite{Arias-Aragon:2025qod}.
It has been argued that 
slight modifications to the operation of this beamline could achieve an estimated intensity of up to $10^{15}$ e$^+$oT per year,
 with a percent-level energy spread and limited hadronic contamination.
In this work we will conservatively assume for the CERN positron beam  $E_B=100\,$GeV and an intensity of $10^{13}$ e$^+$oT per year. Table~\ref{tab:summary} provides a summary of the beam parameters used in this work.
\begin{table}[]
    \centering
    \renewcommand{\arraystretch}{1.2}
    \begin{tabular}{|c|c|c|c|c|}
    \hline
     & $N_{\mathrm{e}^+\mathrm{oT}}$/year & $E_B$ &  $^{74}$W target & $z_D$\\
    \hline
    LNF   &  $10^{18}$ & $450\,$MeV &  $5\,$cm & $3\,$m -- $100\,$m\\
    \hline
    JLab  & $10^{21}$ & $12\,$GeV &  $1\,{\rm cm\, -}\, 5\,$cm & $3\,$m -- $100\,$m\\
    \hline
    CERN  & $10^{13}$ & $100\,$ GeV &  $5\,$cm & $3\,$m -- $100\,$m\\
    \hline
    \end{tabular}
    \caption{Beam parameters for the LNF, JLab and CERN
    experimental setups considered in this work. 
}
    \label{tab:summary}
\end{table}

Backgrounds for resonant $e^+e^-$ production include secondary electrons  detected in coincidence with primary or secondary positrons. This source of noise can be mitigated by measuring their depleted momentum via electromagnetic deflection. 
Another source of background can arise from photons produced via bremsstrahlung in the initial layers of the target. Bremsstrahlung photons carrying a large fraction of the beam energy may convert into $e^+e^-$ pairs within the last millimetres of the target, giving rise to lepton pairs that retain a sizeable fraction of the beam energy.
This background can be substantially suppressed by integrating a veto at the end of the target to ensure that the $e^+e^-$ pairs are created \
outside the target from BSM particle decays (although this might be challenging at LNF and JLab given  the very high beam intensities). 
Finally, a significant background reduction would be obtained by equipping the experiment with a suitable tracking system capable of reconstructing precisely the $e^+e^-$ invariant mass. The precision of the tracking system in reconstructing the original decay vertex will finally be a critical criterium in reducing the pair conversion background. We will argue in the next section that a centimetre-scale precision on displaced vertices would be crucial to probe a large part of the sensitivity gap.  
 
\begin{figure}[h!]
    \centering
    \includegraphics[width=0.85 \textwidth]{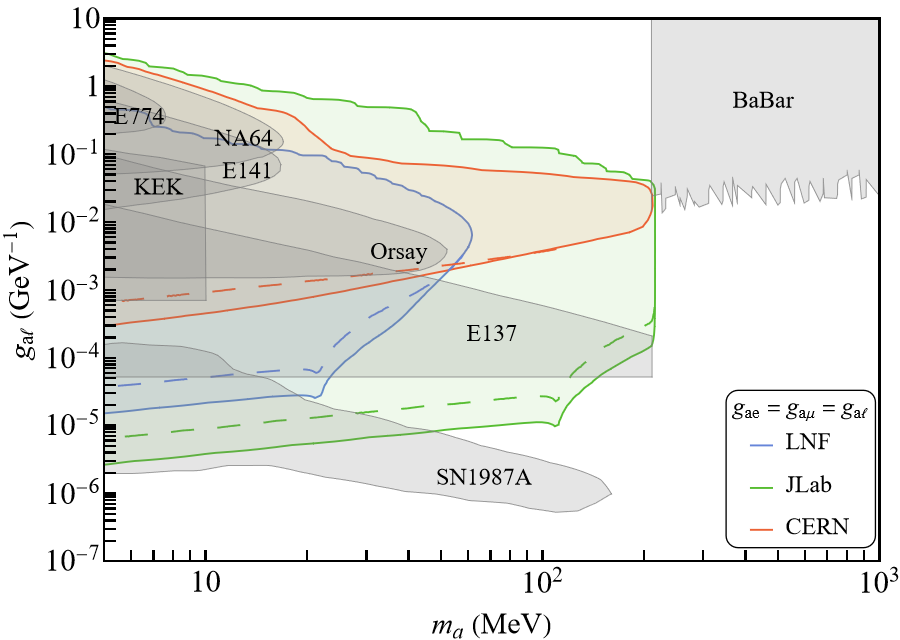}
    \includegraphics[width=0.85 \textwidth]{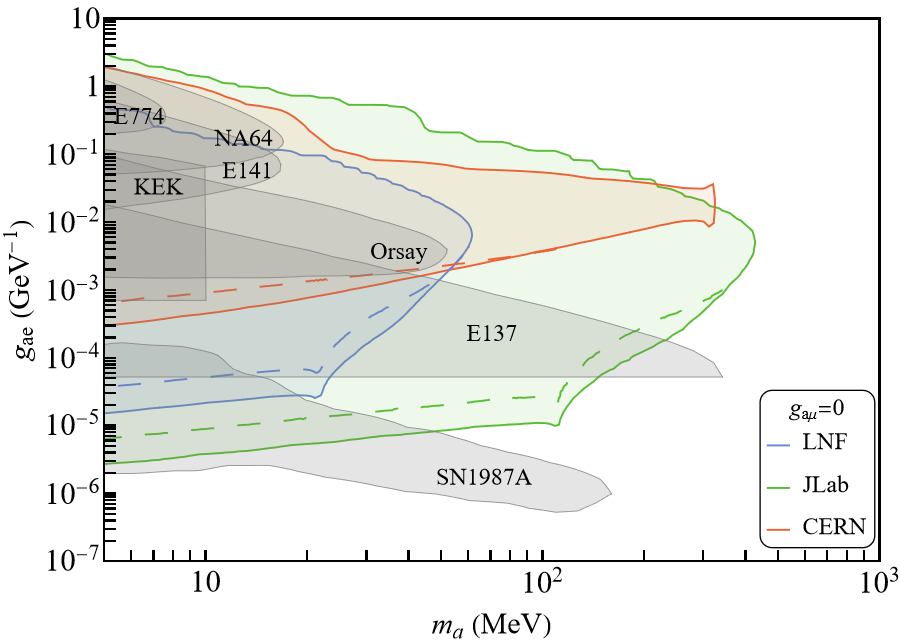}
    \caption{Expected sensitivity for pseudoscalar searches for a $5$ cm tungsten target. In the plot at the top $g_{ae}\approx g_{a\mu} $ has been assumed, while the plot at the bottom assumes $g_{a\mu}\approx 0$. The solid (dashed) lines show the results for a detector set at $100$ m ($3$ m). The regions presently excluded are shaded in gray.
    }
    \label{fig:Pseudo}
\end{figure}

In the following, we assume that appropriate experimental strategies are implemented to suppress to a negligible level backgrounds to the charged di-lepton signal from decays. Figures~\ref{fig:Vec} and~\ref{fig:Pseudo} depict our results on the sensitivity reach achievable with the positron beam parameters listed in Table~\ref{tab:summary}. In these plots, the contours of the region shaded in colour indicate the $90\%$ C.L. sensitivity, corresponding to 2.3 signal events. However, we have verified that the sensitivity regions are not drastically altered by requiring 
100 signal events: the maximum reach in mass is reduced by approximately 25\%,  while the sensitivity to small 
couplings is reduced by about 50\%.

\subsection{Projected Sensitivities to New Vector Particles}

\begin{figure}[t]
    \centering
    \includegraphics[width=0.85 \textwidth]{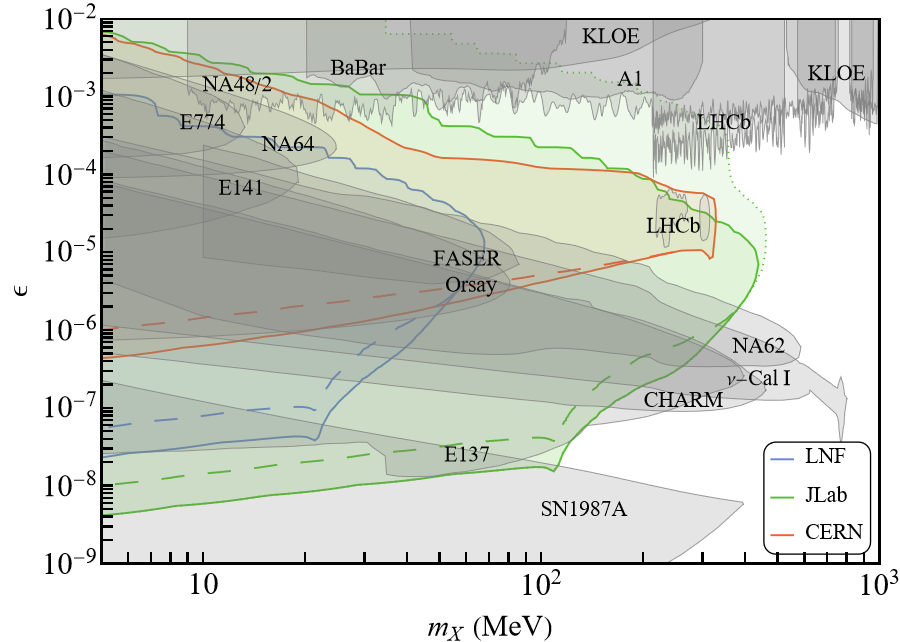}
    \caption{Projected sensitivity for vector searches with a 5\,cm  tungsten target. The solid (dashed) lines show the regions
    that could be excluded with a detector located $100$ m ($3$ m) from the start of the target. The dotted green line corresponds to 
    1\,cm target at JLab. With this setup, it would be possible to fully probe the gap between collider and fixed-target experiments. 
  Regions that have already been excluded by various experiments are shaded in gray~\cite{NA482:2015wmo, BaBar:2014zli,Merkel:2014avp,KLOE-2:2016ydq, LHCb:2019vmc, Bross:1989mp, NA64:2019auh, Riordan:1987aw, FASER:2023tle, Davier:1989wz, Tsai:2019buq, NA62:2023nhs, Blumlein:2011mv, Blumlein:2013cua, Gninenko:2012eq, Bjorken:1988as, Batell:2014mga, Marsicano:2018krp, Chang:2016ntp, Hardy:2016kme}.
    }
    \label{fig:Vec}
\end{figure}

The  projected sensitivities for searches for dark vectors   decaying to $e^+e^-$ and $\mu^+\mu^-$ are shown in Figure~\ref{fig:Vec} assuming a universal coupling $g_{Ve}=g_{V\mu}=\epsilon\cdot e$. 
Current excluded regions are shaded in gray, and include results 
 from NA48/2~\cite{NA482:2015wmo}, BaBar~\cite{BaBar:2014zli}, A1~\cite{Merkel:2014avp}, KLOE~\cite{KLOE-2:2016ydq}, LHCb~\cite{LHCb:2019vmc}, E774~\cite{Bross:1989mp}, NA64~\cite{NA64:2019auh}, E141~\cite{Riordan:1987aw}, FASER~\cite{FASER:2023tle}, ORSAY~\cite{Davier:1989wz}, NA62~\cite{Tsai:2019buq,NA62:2023nhs}, $\nu$-Cal~I~\cite{Blumlein:2011mv,Blumlein:2013cua,Tsai:2019buq}, CHARM~\cite{Gninenko:2012eq,Tsai:2019buq}, E137~\cite{Bjorken:1988as,Batell:2014mga,Marsicano:2018krp}, and SN~\cite{Chang:2016ntp,Hardy:2016kme}.\footnote{Note that
 for the so called {\it protophobic} models,  where couplings to protons and pions are suppressed, the limits from NA48/2 and $\nu$-Cal~I are weaker 
 than those shown in Figure~\ref{fig:Vec}.}

The blue region shows the expected sensitivity achievable with the BTF positron beam at LNF with $E_B = 450\,$MeV, a thick target of $5\,$cm, and an intensity of $10^{18}$ e$^+$/oT. 
The solid (dashed) curve corresponds to a detector positioned at a distance of $100\,$m ($3\,$m) from the start of the target. With a far detector, the bounds at small couplings are  strengthened due to the increased number of BSM vectors decaying before reaching the detector.

The CEBAFe+ beam at JLab with $E_B = 12\,$GeV, an intensity of $10^{21}$ e$^+$/oT, and a target size of $5\,$cm could probe the green region. 
The solid and dashed curves are respectively for a 
target-detector distance of 100\,m and 3\,m. 
For JLab, we also show 
 with the green dotted line the sensitivity that could be obtained with a $1\,$cm target.  The 
  enhanced sensitivity at large couplings, driven by the increased number of BSM vectors decaying outside the target, enables full exploration of the parameter space between collider and fixed-target experiment limits. We have  verified that 
  this remains true even when  we require 100 signal events.
 
 Finally, the orange region represents the possible reach achievable with the 
 H4 positron beamline  at CERN with $E_B = 100\,$GeV and an intensity of $10^{13}$ e$^+$/oT. The solid and dashed lines refer respectively to a
 target-detector distance of 100\,m and 3\,m. 
 
 For all the three cases, the upper boundary of the sensitivity region is determined  by the target size, as above this boundary the large couplings imply that dark particles decay within the target. Conversely, the lower boundary at small couplings is determined  by the production rates, as well as by the distance between the target and the detector. 
Although fewer dark vectors are produced at smaller couplings, placing detectors farther away would ensure that all of them will have time 
to decay into detectable charged leptons. This helps to probe a larger portion of the parameter space.

The results displayed in Figure~\ref{fig:Vec}  emphasize the role of higher beam intensities in increasing the accessible mass region.
We observe that the one-order-of-magnitude lower beam energy at JLab, compared to CERN, is more than compensated by its significantly higher beam intensity. This enables a substantial number of collisions with  atomic electrons in the 
high-momentum tail of the distribution, leading to a considerable enhancement of the c.m. energy of the collisions.

Finally, although we only present results for spin-one vector particles, we have verified that, to an excellent approximation, the mass/coupling exclusion region shown in Figure~\ref{fig:Vec} also applies to BSM axial-vector particles
 $A_\mu$ coupled to leptons via the interaction Lagrangian $\mathcal{L}_A = \sum_\ell g_{A\ell} A_\mu \bar{\ell} \gamma^\mu \gamma_5\ell$.

\subsection{Projected Sensitivities to New Pseudoscalar Particles}

In Figure~\ref{fig:Pseudo} we present the sensitivity reach  
for searches of  BSM pseudoscalars resonantly produced in positron  annihilation on atomic electrons using the LNF, JLab and CERN positron beams.
Other  proposals for dedicated pseudoscalar searches that are complementary to 
 our approach have been put forth, see for example 
 Refs.~\cite{Bertuzzo:2022fcm,Capozzi:2023ffu,GrillidiCortona:2022kbq}.
Existing experimental bounds (all of which were derived without accounting for electron motion effects) are represented by the gray-shaded regions. They were obtained by NA64~\cite{NA64:2021aiq}, E141~\cite{Riordan:1987aw}, E774~\cite{Bross:1989mp}, BaBar~\cite{
BaBar:2016sci,Bauer:2017ris}, KEK~\cite{Konaka:1986cb}, E137~\cite{Bjorken:1988as}, Orsay~\cite{Davier:1989wz} and from limits on supernova SN1987A extra energy loss~\cite{Carenza:2021pcm,Fiorillo:2025sln}.
Note that besides the experimental limits shown in the plots, 
other limits  derived in theoretical studies exist~\cite{Batell:2009jf,Freytsis:2009ct,Dolan:2014ska,Choi:2017gpf,
Dobrich:2018jyi,MartinCamalich:2020dfe,Carmona:2021seb,Guerrera:2021yss,
Bauer:2021mvw,Altmannshofer:2022ckw}.
Since they often apply to specific constructions or involve model 
dependencies, they have not been included in the figures.  

In the top panel we show the results for a pseudoscalar that couples 
to electrons and muons with similar strength $g_{ae}\approx g_{a\mu}$ (for the numerical computation of the results, we set $g_{ae}= g_{a\mu}$ but small deviations from this would not alter the limits in any noticeable way).
The sharp drop in the sensitivity to $g_{ae}$ 
at $m_a \simeq 2m_\mu$, clearly visible in the JLab green curve,  arises from the opening of the muonic decay channel. 
This channel is enhanced by a factor $(m_\mu/m_e)^2$  
relative to the electron channel, resulting in  ${\rm Br}(a\to e^+e^-) \approx 0$. 
The same effect is visible in the E137 excluded region, and to a lesser extent 
in the CERN orange region. 
Conversely, the  BaBar  excluded  region begins at 
$m_a \simeq 2m_\mu$ and extends to larger masses. This is because 
this limit, derived in Ref.~\cite{Bauer:2017ris} by reinterpreting 
the BaBar limit on vector particles~\cite{BaBar:2016sci}, is obtained from 
the process $e^+ e^-\to \mu^+\mu^- X$ followed by $X\to \mu^+\mu^-$, for which phase space closes below $m_a \simeq 2\,m_\mu$. 
In this and the following plot, we do not display the results for JLab with a 1 cm target. While sensitivity would improve in this case as well,  it would not close any critical gap in the parameter space.
Finally, let us mention for completeness that if $g_{ae}\approx g_{a\mu}$ is assumed, the SN1987A limit could become significantly stronger than what 
shown in the plot~\cite{Bollig:2020xdr,Croon:2020lrf,Caputo:2021rux}. 
This is because  pseudoscalar production via interactions with 
 muons in the protoneutron star would be greately enhanced due to 
 the much larger coupling strength.

The bottom panel depicts the sensitivity to a pseudoscalar that couples mainly to electrons, while $g_{a\mu}\approx 0$. In this case  the BaBar limit disappears, while the JLab, CERN and E137 exclusion regions now extend up to masses  well beyond $m_a \simeq 2 m_\mu$.

Finally, we have verified that  also in the case of spin zero particles, to an excellent approximation, the  mass/coupling exclusion regions shown in the two panels of  
Figure~\ref{fig:Pseudo} remain valid for a BSM scalar particle $\sigma$,  
 provided that the scalar-lepton coupling is normalised 
 as in \eq{eq:Pseudo},  i.e. $\mathscr{L}_{\phi} = 
-\sum_\ell m_\ell\, g_{\phi\ell}\, \phi\,\bar{\ell}\ell$. 

\section{Conclusions}
Our study focussed on the resonant production  of BSM particles 
in positron annihilation on fixed-target atomic electrons. We have 
considered the positron beams of three experimental facilities 
-- LNF, JLab and CERN --  impinging on a tungsten target.
We have demonstrated that for all three experimental setups, which feature significantly different beam energies and intensities, a reliable assessment of the mass reach in searches for new particles must account for the effects of atomic electron momentum distribution.
We have identified an intriguing interplay between higher beam energy and higher beam intensity. The latter can more than compensate for the former in achieving greater centre-of-mass collision energy, as high-intensity beams increase the likelihood of collisions with atomic electrons in the high-energy tail of the distribution, where electron density is significantly suppressed. 
For this reason, the 12 GeV Ce$^+$BAF positron beam at JLab, with a forecasted intensity eight orders of magnitude higher than the 100 GeV H4 positron beam at CERN, can probe a significantly larger parameter space. This enhanced sensitivity extends not only to couplings two orders of magnitude smaller but also to new particle masses up to twice as large as those accessible at CERN.
Finally, while we have considered $^{74}$W as the target material, which is a common choice in fixed-target experiments, the effects of atomic electron momenta will be significantly enhanced in higher-$Z$ materials such as $^{84}$Pb, and even more so in $^{90}$Th or $^{92}$U. Therefore, their suitability as target materials warrants dedicated exploration.

\section*{Acknowledgement}
We are deeply grateful to Luc Darmé for his fundamental contributions to our understanding of atomic electron momentum effects.
We thank  Lau Gatignon and Johannes Bernhard for providing  
detailed information on the CERN NA beams,
and Eric Voutier for sharing details on the CEBAF positron beam parameters. 
We acknowledge conversations with  M. Raggi and P. Valente. 
F.A.A., G.G.d.C. and E.N. are supported in part by the INFN ``Iniziativa Specifica" Theoretical Astroparticle Physics (TAsP). F.A.A. received additional support from an INFN Cabibbo Fellowship, call 2022.
G.G.d.C. acknowledges LNF for hospitality at various stages of this work. The work of E.N. is  supported  by the Estonian Research Council grant PRG1884.   Partial support from the CoE grant TK202 “Foundations of the Universe” and from the CERN and ESA Science Consortium of Estonia, grants RVTT3 and RVTT7, and  
 from the COST (European Cooperation in Science and Technology) Action COSMIC WISPers CA21106 are also acknowledged.

\footnotesize

\bibliographystyle{elsarticle-num}
\bibliography{biblio}{}

\end{document}